\documentclass[11pt]{article}
\usepackage{amsmath,amssymb,color}
\usepackage{graphicx}

\usepackage{amsfonts}

\newcommand{\hoch}[1]{$\, ^{#1}$}

\newcommand{\be}{\begin{equation}}
\newcommand{\ee}{\end{equation}}
\newcommand{\bea}{\begin{eqnarray}}
\newcommand{\eea}{\end{eqnarray}}

\newcommand{\mt}[1]{\textrm{\tiny #1}}
\def\nc {N_\mt{c}}

\newcommand{\uh}{u_\mt{H}}

\newcommand{\cf}{{\cal F}}
\newcommand{\cb}{{\cal B}}
\newcommand{\ch}{{\cal H}}

\def\ft#1#2{{\textstyle{\frac{\scriptstyle #1}{\scriptstyle #2} } }}

\def\0{{\sst{(0)}}}
\def\1{{\sst{(1)}}}
\def\2{{\sst{(2)}}}
\def\3{{\sst{(3)}}}
\def\4{{\sst{(4)}}}
\def\5{{\sst{(5)}}}
\def\6{{\sst{(6)}}}
\def\7{{\sst{(7)}}}
\def\8{{\sst{(8)}}}
\def\sst#1{{\scriptscriptstyle #1}}

\begin{document}

\begin{flushright}
\hfill{ \
SHUITP-14-01\ \ \ \ }
\end{flushright}

\vspace{25pt}
\begin{center}
{\large {\bf Anisotropic plasma with a chemical potential and scheme-independent instabilities}}

\vspace{15pt}

   Long Cheng\hoch{1}, Xian-Hui Ge\hoch{1} , Sang-Jin Sin\hoch{2}

\vspace{10pt}

\hoch{1}{\it Department of Physics\\
Shanghai University, Shanghai 200444, China}

\vspace{10pt}

\hoch{2}{\it Department of Physics, Hanyang University\\
Seoul 133-791, Korea}

\vspace{10pt}

\vspace{10pt}

\vspace{40pt}

\underline{ABSTRACT}
\end{center}
Generically, the black brane solution with planar horizons is thermodynamically stable. We find a counter-example to this statement by demonstrating that an anisotropic black
brane is unstable.
  We present a charged black brane solution dual to a spatially anisotropic finite temperature $\mathcal{N}=4$ super Yang-Mills plasma at finite $U(1)$ chemical potential.
  This static and regular solution is obtained both numerically and analytically. We uncover rich thermodynamic phase structures for this system by considering the cases when the anisotropy
  constant ``a" takes real and imaginary values, respectively.  In the case $a^2>0$, the phase structure of this anisotropic black brane is similar to that of Schwarzschild-AdS black hole with $S^3$ horizon topology, yielding a thermodynamical instability at smaller horizon radii. For the condition $a^2\leq 0$, the thermodynamics is dominated by the black brane phase for all temperatures.


\vspace{15pt}

\thispagestyle{empty}





\newpage
\section{Introduction}
The AdS/CFT correspondence provides a powerful tool in studying the strongly coupled problems of quantum field theory, ranging from nuclear physics to condensed matter theory\cite{duality,duality2}.
This correspondence states the equivalence between type IIB superstring theory in $AdS_5\times S^5$ and $\mathcal N = 4$ super Yang-Mills (SYM) gauge theory on the $4-$dimensional boundary of $AdS_5$. From gravitational theory on the asymptotically anti-de Sitter view point, we are able to gain profound insights for such strongly coupled field theory. It is thus very crucial
to search for the generic asymptotically AdS gravitational solutions, which are dual to interesting phase in the field theory side. The most well-known black brane solutions are the homogeneous and isotropic Schwarzschild-AdS black brane solution and Reissner-Nordstr$\ddot{o}$m-AdS (RN-AdS) solution. The charged black brane solutions are particularly useful to study quark-gluon plasma (QGP)\cite{son}, superconductivity and superfluidity, Fermi surfaces and non-Fermi liquids in condensed matter system\cite{gubser,horowitz,hartnoll,herzog,liu}. Generally, $U(1)$ gauge symmetries in the bulk correspond to conserved number operators in the dual field theory.
  The gauge field in the AdS space couples to a CFT current $J_{\mu}$ and the CFT states thus containing a plasma of charged quanta.

It is well-known that there are many strongly coupled  systems which do not satisfy homogeneity and isotropy spontaneously. For example, some systems may have anisotropic Fermi-surface because of the atomic lattice effects and the QGP is anisotropic in a short time after creation . Therefore, the studies on anisotropic and inhomogeneous black brane solutions and their holographic applications  have attracted more attention \cite{mateos,anisotropic,inh}.

In this paper, we will present a charged and spatially anisotropic black brane solution. The neutral anisotropic black brane solution was obtained by  Mateos and Trancanelli in their seminal papers \cite{mateos} and its applications in QCD was discussed.
One motivation comes from the fact that the QGP created in RHIC is  not only anisotropic but also charged.  In the QGP produced in RHIC, the escaped quark is surrounded by high density quark fluid liberated from the heavy ions. Under such conditions, the baryon density  of the QGP and the overall $U(1)$ gauge field is relevant. Unlike chargeless case, the introduction of the $U(1)$ gauge field breaks the $SO(6)$ symmetry and thus leads to the excitations of the Kaluza-Klein modes. Another motivation comes from the applications of the anisotropic black brane solutions to condensed matter physics, since the many-body system at a finite $U(1)$ charge density  corresponds to the charged black holes in the AdS peace.

   We will consider the case the anisotropy is introduced through deforming the SYM theory by a $\theta$-parameter of the form $\theta\propto z$, which acts
   as an isotropy-breaking external source that forces the system into an anisotropic equilibrium state \cite{mateos}. The $\theta$-parameter is dual to the type IIB axion $\chi$ with the form $\chi=az$. The constant $a$ has dimensions of mass and is a measure of the anisotropy. From the five-dimensional theory viewpoint, the anisotropy can be interpreted
   as a non-zero number of dissolved D7-brane wrapped on $S^5$, extending along the $xy$-direction and distributed along the $z$-direction with density $n_{D7}$ \cite{mateos}. So $a$ can be regarded as ``charge density" and should not be imaginary-valued.

   However,  we will consider both cases with $a^2> 0$ and $a^2< 0$ which have different thermodynamic properties, although imaginary axion field might be unphysical in type IIB supergravity theory.
   If the axion field merely plays the role of providing the appropriate source to support a spatially anisotropic spacetime, then the imaginary-valued $a$ could be acceptable. Furthermore, we will see later that imaginary $a$ can be understood as a consequence of
   the tachyon condensate of the dilaton field. A careful analysis in the following will disclose that the anisotropic black brane solution corresponding to $a^2>0$ is actually a ``prolate" version
   of the solution because it has a $z$-axis longer than the $x$ and $y$-axes (i.e. $\ch(\uh)>1$), while the ``oblate" version of the anisotropic black brane solution requires $a^2<0$ (i.e.
   $\ch(\uh)<1$) and thus the $z$-axis is shorter than the $x$ and $y$-axes. As what we will uncover, the ``prolate" black brane solution suffers thermodynamic instabilities, similar to those of the Schwarzschild-AdS with a spherical horizon, but the ``oblate" solution is stable.


\section{Numerical Solution}
The charged anisotropic black brane solution can be derived from the effective action after $S^5$ reduction of type IIB supergravity\cite{lvhx,cgs}. In Einstein frame, the type IIB supergravity Lagrangian which have been truncated out NS-NS and R-R 2-form potentials is
\bea
{\cal L}= \hat{R}*1-\frac{1}{2}d\hat{\phi}\wedge*d\hat{\phi}-\frac{1}{2}e^{2\hat{\phi}} \hat{F}_1\wedge*\hat{F}_1-\frac{1}{4}\hat{F}_5\wedge*\hat{F}_5,
\label{10daction}
\eea
where $\hat{\phi}$ and $\hat{F}_1=d\hat{\chi}$ are the dilaton and the axion field-strength in ten-dimensions respectively.
The 5-form field $\hat{F}_5$ should satisfy the self-duality condition and be imposed at the level of equations of motion.
The theory can be reduced on to minimal supergravity and the corresponding five-dimensional axion-dilaton-Maxwell-gravity action is given by
\be\label{5action}
S=\frac{1}{2\kappa^2}\int d^5x \sqrt{-g} \Big( R +12-\ft12(\partial\phi)^2 - \ft12 e^{2\phi} (\partial \chi)^2- \ft14 F_{\mu\nu}F^{\mu\nu}\Big) +S_{GH}\,,
\ee
where we have set the AdS radius $L=1$, $\kappa^2=4\pi^2/N^2_c$ and $S_{GH}$ is the Gibbons-Hawking boundary term.


In order to obtain an anisotropic D3-brane with an asymmetry between the $xy-$ and $z-$ directions, we assume the Einstein-frame metric takes the form
\bea
&&ds_5^2 =  \frac{e^{-\frac{1}{2}\phi}}{u^2}\left( -\cf \cb\, dt^2+dx^2+dy^2+ \ch dz^2 +\frac{ du^2}{\cf}\right).  \label{10dmetric}\\
&& A=A_t(u)dt,~~~{\rm and }~~\chi=az.
\eea
The functions $\phi$, $\cf$, $\cb$ and $\ch=e^{-\phi}$ depend only on the radial coordinate $u$, which we solved numerically \cite{cgs}. The electric potential $A_t$  can be obtained via $A_{t}(u)=-\int^u_{\uh}du Q\sqrt{\cb}e^{\frac{3}{4}\phi}u $ from the Maxwell equations, where  $Q$ is an integral constant related to the charge. The horizon  locates at $u=\uh\equiv 1/r_{H}$ with $\cf(u_H)=0$ and the boundary is at $u=0$ where $\cf=\cb=\ch=1$. The asymptotic $AdS_5$ boundary condition requires the boundary condition $\phi(0)=0$.
The Hawking temperature is given by $T  =  -\frac{\cf'(\uh) \sqrt{\cb_\mt{H}}}{4\pi} \,$ through the Euclidean method.

\begin{figure}[htbp]
 \begin{minipage}{1\hsize}
\begin{center}
\includegraphics*[scale=0.50] {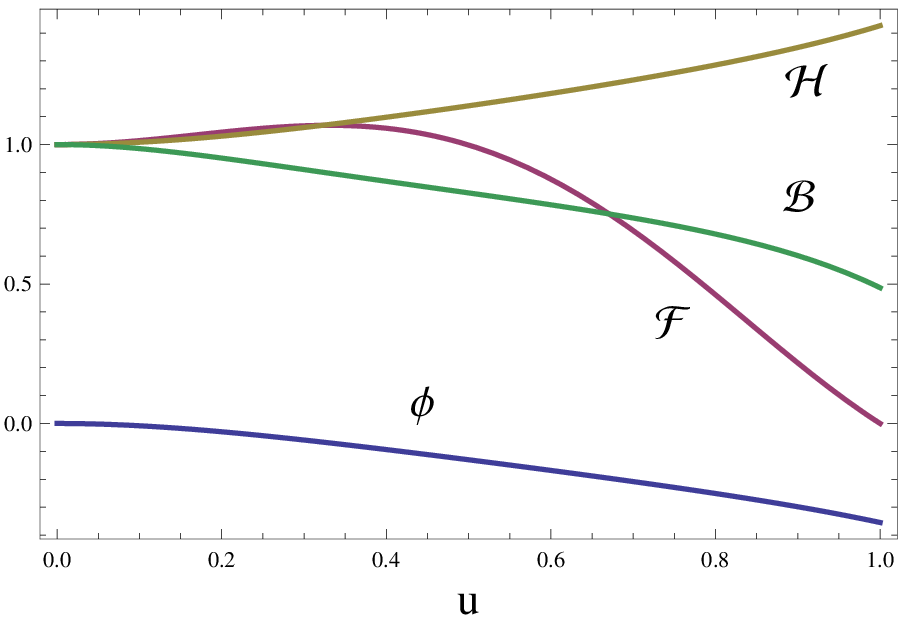}
\includegraphics*[scale=0.50]{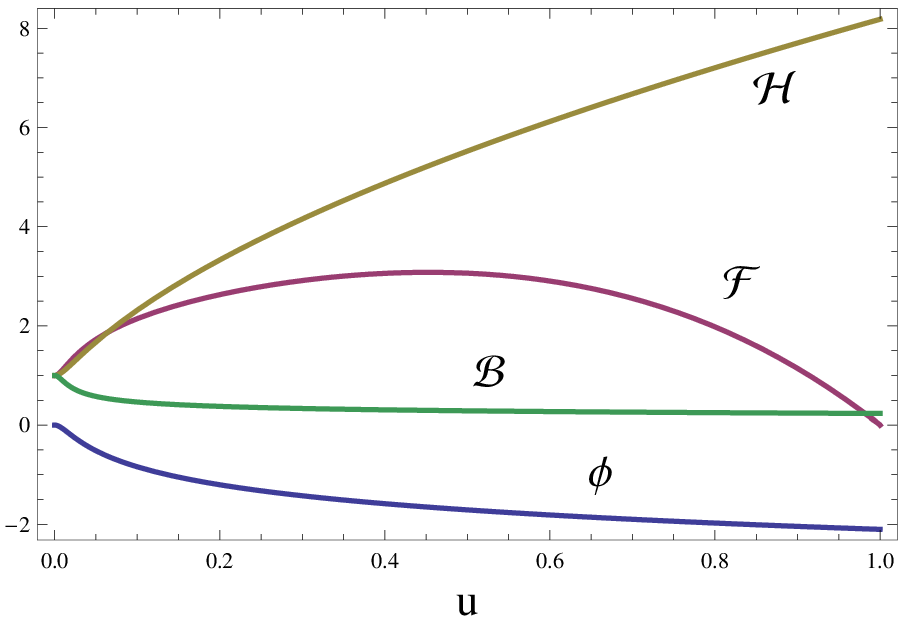}
\includegraphics*[scale=0.50] {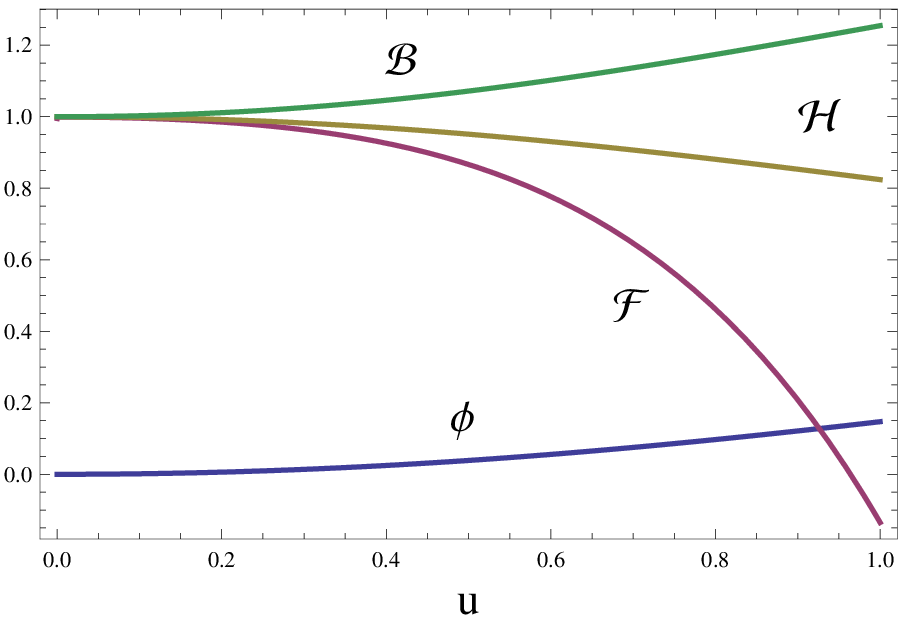}
\end{center}
\caption{(Color online.)The metric functions for $a=1.86$, $Q=6.23$(left), $a=64.06$, $Q=9.76$ (middle) and  $a=1.2i$, $Q=1/10$(right), with $u_H=1$.} \label{metric}
\end{minipage}
\end{figure}

\begin{figure}[htbp]
 \begin{minipage}{1\hsize}
\begin{center}
\includegraphics*[scale=0.40] {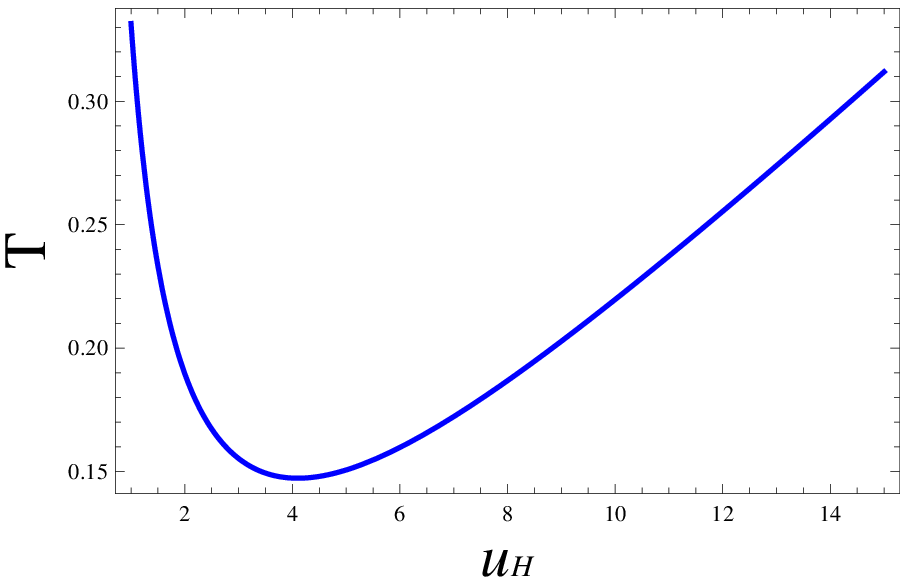}
\includegraphics*[scale=0.40] {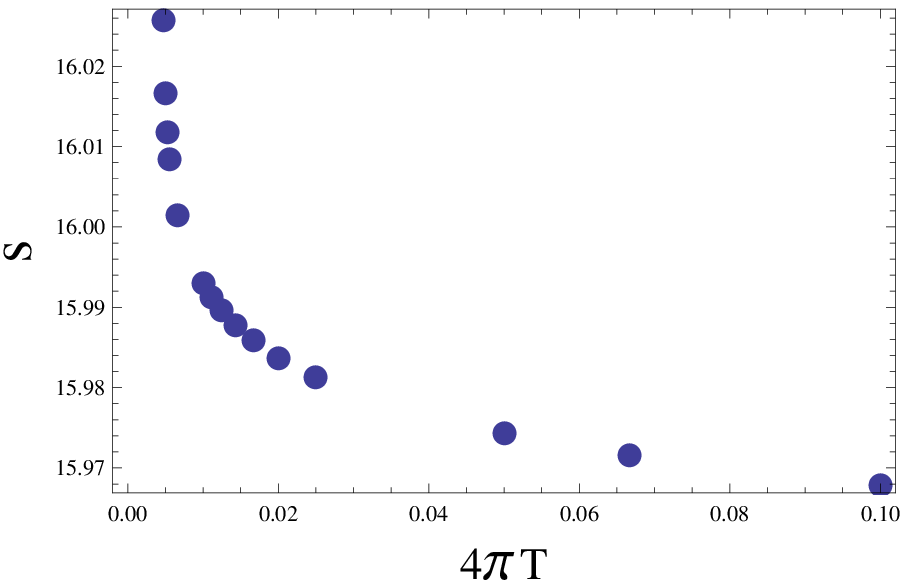}
\includegraphics*[scale=0.40] {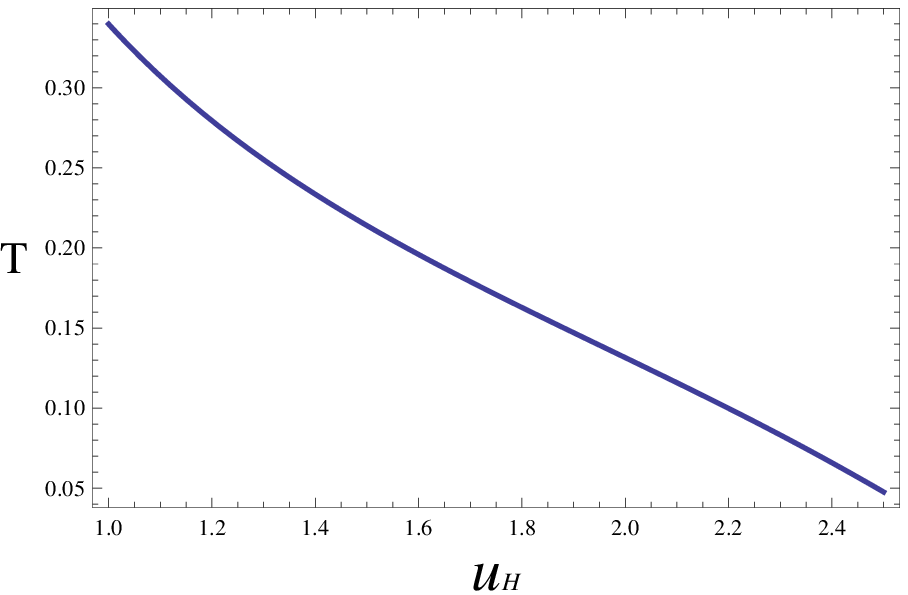}
\includegraphics*[scale=0.40] {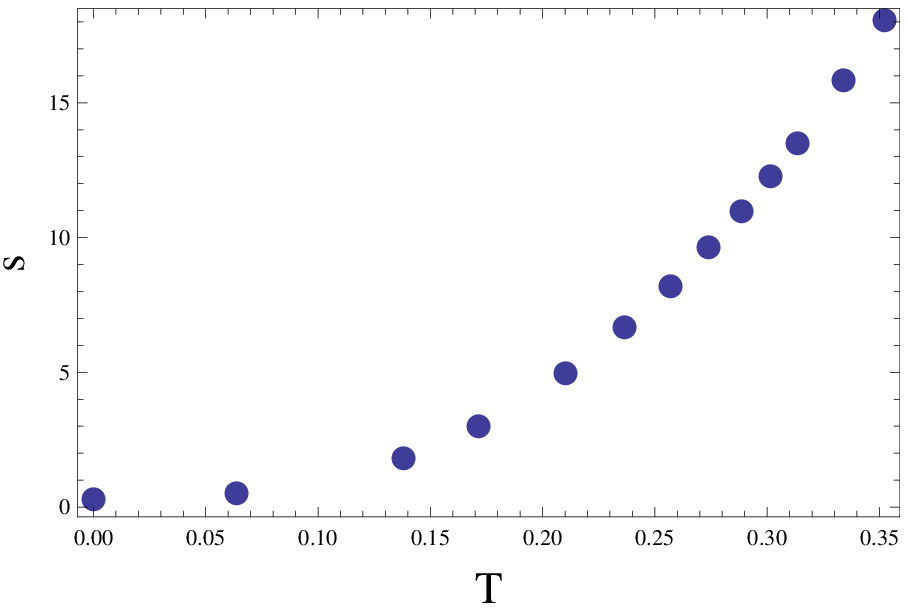}
\end{center}
\caption{(Color online.)  The Hawking temperature and the entropy density as  functions of the inverse horizon radius $\uh$, and the temperature, respectively.
For the first two graphs correspond to ``prolate" black brane solution with $a^2>0$, where the initial conditions is chosen as $\phi_H=-0.22$, $a=1.46$ and $Q=0.01$. The last two graphs plot the ``oblate" version of the black brane solution with $a^2<0$, where the initial conditions are chosen as $\phi_H=0$, $a=1.2i$ and $Q=0.2$.} \label{inverseT}
\end{minipage}
\end{figure}
Fig.\ref{metric} depicts the metric functions corresponding to different initial conditions. The first two plots in Fig.\ref{metric} reflect that the profile for $\cb$ is seriously suppressed at the horizon as the charge $Q$ increases. The Hawking temperature depends strongly on $Q$ and the anisotropy $a$ is sensitive to the initial condition $\phi(\uh)$.  We can also see from Fig.\ref{metric} (right) that even the anisotropy constant $a$ takes imaginary value, the black brane solution is still regular. Note that the metric functions $\ch(\uh)>1$ and $\cb(\uh)<1$ for $a^2>0$, corresponding to the ``prolate" solution, but $\ch(\uh)<1$, $\cb(\uh)>1$ for $a^2<0$, corresponding to the ``oblate" solution.

Note that the temperature is determined by the inverse horizon radius $\uh=1/r_H$ and the charge $Q$.  As can be seen from Fig.\ref{inverseT} (left), for a given temperature there are two branches of allowed black brane solutions, a branch with larger radii and one with smaller.
 This intriguing behavior is similar to the case of Schwarzschild-AdS black holes with a spherically horizon \cite{HP}. The smaller branch of the black brane is unstable with negative specific heat.

 It is well-known that for black brane solutions with horizon topology $R^3$, there is only one branch of black brane solutions and the free energy is negative definite, so that the black brane structure is trivial and the thermodynamics is dominated by the black brane for all temperatures \cite{em}. However, the anisotropic black brane solution obtained here provides a  counter example to the above statement. We notice that even in the absence of $U(1)$ gauge field, two branches of black brane solution still exist, reflecting that it is mainly caused by the anisotropy.
 This behavior was not noticed in \cite{mateos} and all the numerical computation was carried out at the stable black brane branch.

 As to the ``oblate" solution with $a^2<0$, the behavior of the solution differs sharply from the real anisotropy situation, which is qualitatively the same as the planar black brane case \cite{em}: In that situation, there is only one stable branch of black brane solution and the thermodynamics is dominated by this solution for all temperatures (see Fig.\ref{inverseT} (third)). The entropy density decreases as temperature goes down so that the specific heat is positive $c_{\rho}=T(\partial s/\partial T)_{\rho}>0$.

 \textbf{Extremal limit} As shown in Fig.\ref{inverseT} (left), for the ``prolate" solution with  anisotropy parameter $a^2>0$, the temperature $T$ cannot reach zero and thus this charged anisotropic configuration has no extremal limit, which is consistent with \cite{maeda}.  This is further supported by Fig.\ref{inverseT} (second), which plots the entropy density as a function of the temperature. Furthermore, Fig.\ref{inverseT} (second) shows that the entropy density increases as the temperature goes down, which implies an instability of the black brane, since the heat capacity is then negative. This result further support our previous argument  that the ``prolate" black brane behaves like the Schwarzschild-AdS black hole with the spherical horizon.

 In the case of $a^2 <0$, it is clear that there exists the extremal black brane solution as shown in Fig.\ref{inverseT} (third). Numerical computation implies that even close to zero temperature, the system prefers to be dominated by black brane with non-zero entropy. We will provide a consistent check on the above arguments by using the following analytic study.


 \section{Analytic solution in small-anisotropy limit}

 The analytic  black brane solution in small anisotropy limit is obtained to the leading order in $a$ by perturbating the RN-AdS black brane solution. We hope the analytic solution can help us pick out more physics in a straightforward way.
 The functions $\cf$, $\cb$ and $\ch$ can be expressed as
 \bea
 &&\cf=1-\bigg(\frac{u}{u_H}\bigg)^4+\bigg[\bigg(\frac{u}{u_H}\bigg)^6-\bigg(\frac{u}{u_H}\bigg)^4\bigg]q^2+a^2 \cf_2(u)+\mathcal{O}(a^4),\\
 &&\cb=1+a^2 \cb_2(u)+\mathcal{O}(a^4),\\
 &&\ch=e^{-\phi(u)}, {~~~\rm with ~~~}  \phi(u)=a^2 \phi_2(u)+\mathcal{O}(a^4),
 \eea
 where
 \bea
 \cf_2(u)&=&\frac{1}{24\sqrt{1+4q^2}\uh^4}\bigg\{3(-4q^2u^6+\uh^6)\log\left(\frac{(1+\sqrt{1+4q^2})u^2+2\uh^2}{(1-\sqrt{1+4q^2})u^2+2\uh^2}\right) \nonumber\\[0.7mm]
 &+&u^2\uh^2\bigg[8\sqrt{1+4q^2}(-u^2+\uh^2)+u^2\big(3\log\left(-2-2q^2+2\sqrt{1+4q^2})\right)\nonumber\\[0.7mm]
 &+&5(-2+q^2)\log\left(-1+2q^2+\sqrt{1+4q^2}\right)-12q^2\log\left(-2-2q^2+2\sqrt{1+4q^2}\right)\nonumber\\[0.7mm]
 &+&7(1+q^2)\Big(\log\left((-1+2q^2-\sqrt{1+4q^2})(2q^2u^2+(-1+\sqrt{1+4q^2})\uh^2)\right)\nonumber\\[0.7mm]
 &-&\log\left(2q^2u^2-(1+\sqrt{1+4q^2})\uh^2\right)\Big)\bigg]\bigg\},\nonumber\\[1.7mm]
 \phi_2(u)&=&\frac{\uh^2}{4\sqrt{1+4q^2}}\log\left(\frac{(1+\sqrt{1+4q^2})u^2+2\uh^2}{(1-\sqrt{1+4q^2})u^2+2\uh^2}\right),\nonumber\\[1.7mm]
 \cb_2(u)&=&\frac{\uh^2}{24}\left(\frac{10u^2\uh^2}{q^2u^4-u^2\uh^2-\uh^4}+\frac{1}{\sqrt{1+4q^2}}\log\left(\frac{(1+\sqrt{1+4q^2})u^2+2\uh^2}{(1-\sqrt{1+4q^2})u^2+2\uh^2}\right)\right).
 \eea
The parameter $q$ denotes the dimensionless charge parameter with $q=\frac{\uh^3 Q}{2\sqrt{3}}$ and the physical range of $q^2$ is $0\leq q^2<2$. The electrical potential is given by
\bea
A_t&=&\frac{q}{8\uh^3\sqrt{3+12q^2}}\Big(24\sqrt{1+4q^2}(\uh^2-u^2)+5a^2\uh^2\Big[\uh^2\log\left(\frac{3-\sqrt{1+4q^2}}{3+\sqrt{1+4q^2}}\right)\nonumber\\[0.3mm]
 &&+u^2\log\left(\frac{(1+\sqrt{1+4q^2}u^2)+2\uh^2}{(1-\sqrt{1+4q^2})u^2+2\uh^2}\right)\Big]\Big)+\mathcal{O}(a^4).
\eea

 \begin{figure}[htbp]
 \begin{minipage}{1\hsize}
\begin{center}
\includegraphics*[scale=0.5] {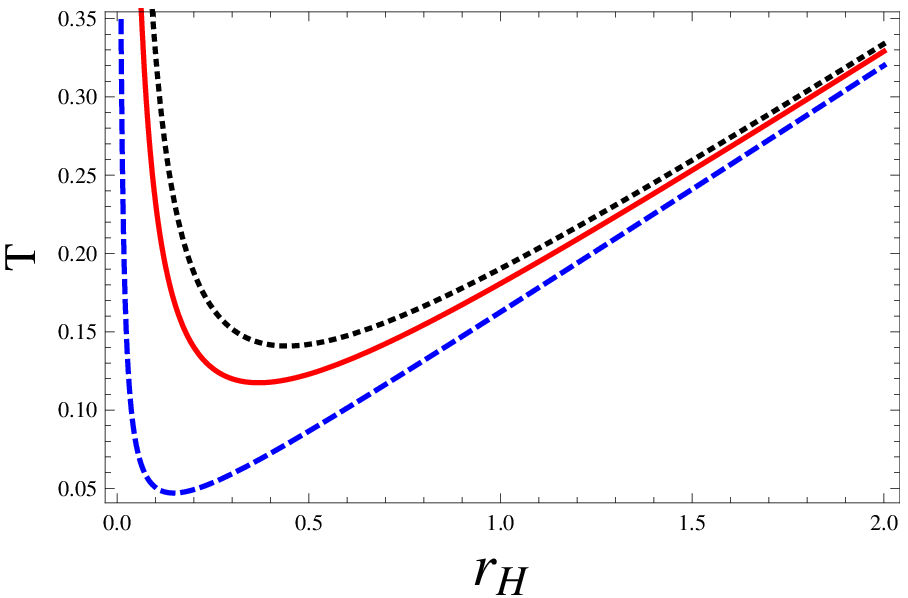}
\includegraphics*[scale=0.52] {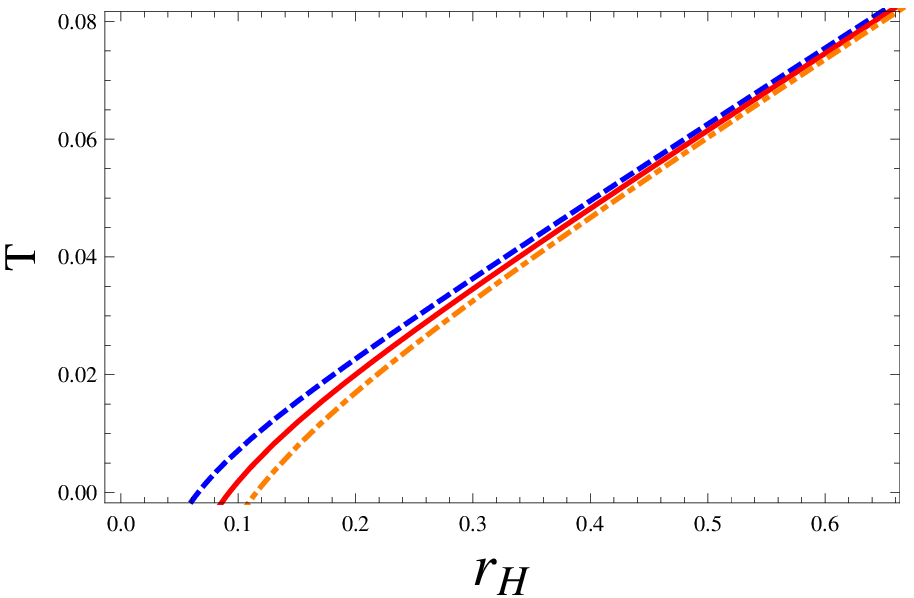}
\includegraphics*[scale=0.5] {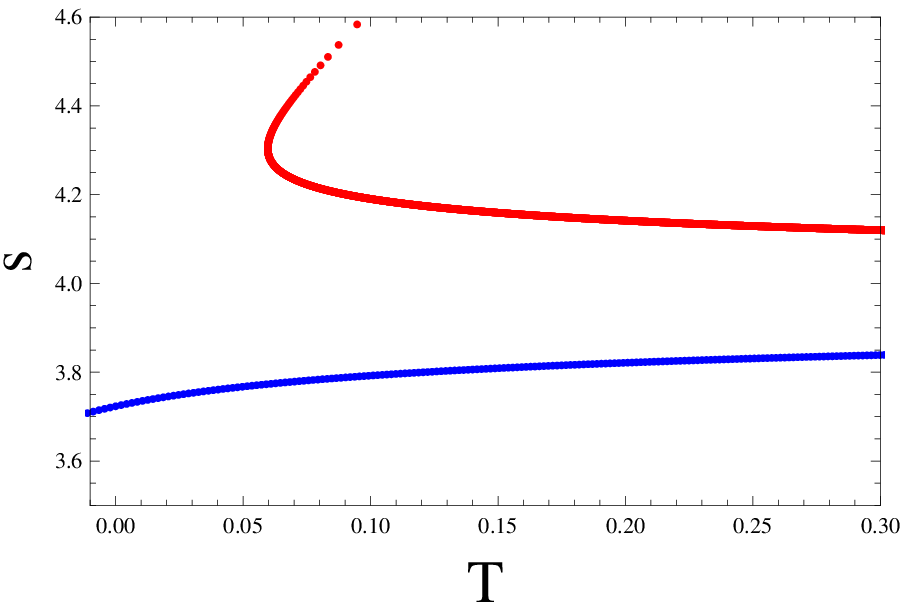}
\end{center}
\caption{(Color online.) (Left)Temperature as a function of the horizon radius for $a=1/5$(blue), $a=1/2$(red) and $a=3/5$ (black), where $N_c=1$ and $q=1$. (Middle) Temperature vs $r_H$ for $a=i/14$(blue), $a=i/10$(red) and $a=i/8$(orange), where $N_c=1$ and $q=1.2$. (Right) Entropy density as a function of temperature for ``prolate" solution (red) and ``oblate" solution (blue), respectively. } \label{solx}
\end{minipage}
\end{figure}
By using the Euclidean method, we can easily obtain the Hawking temperature
 \bea\label{tema}
T&=&-\frac{\cf'(\uh) \sqrt{\cb_\mt{H}}}{4\pi}\nonumber\\
&=&\frac{2-q^2}{2\pi \uh}+\frac{\uh\left(-4\sqrt{1+4q^2}+5(2+5q^2)\log\big(\frac{3+\sqrt{1+4q^2}}{3-\sqrt{1+4q^2}}\big)\right)}{96\pi\sqrt{1+4q^2}}a^2+\mathcal{O}(a^4).
 \eea
 The entropy density derived from the Bekenstein-Hawking formula is written as
 \bea
 s &=& \frac{A_\mt{H}}{4 G V_3}=\frac{\nc^2 e^{-\frac{5}{4}\phi_H}}{2\pi \uh^3}\nonumber\\
 &=&\frac{N^2_c}{2\pi \uh^3}+\frac{5\nc^2\log\big(\frac{3+\sqrt{1+4q^2}}{3-\sqrt{1+4q^2}}\big)}{32\pi\sqrt{1+4q^2}\uh}a^2+\mathcal{O}(a^4),
 \eea
 where $V_3$ is the volume of the black hole horizon.
 The chemical potential is obtained as
 \bea\label{mu1}
\mu
 &=&\frac{q}{8\sqrt{3}\uh}\left(24+\frac{5\uh^2\log(\frac{3-\sqrt{1+4q^2}}{3+\sqrt{1+4q^2}})}{\sqrt{1+4q^2}}a^2\right)+\mathcal{O}(a^4).
\eea

 We notice that for ``prolate" solution with  $a^2>0$ at finite temperature, the horizon radius $r_{H}$ and the entropy density  of the anisotropic black brane are greater than that of isotropic RN-AdS black brane.
 Thus the chemical potential is less than that of isotropic RN-AdS black brane $\mu^0$. The temperature as a function of the horizon radius is shown in Fig.\ref{solx} (left and middle) for the cases of real and imaginary
 anisotropy parameter $a$ respectively. The first graph in Fig.\ref{solx} shows that there are two branches for the small-anisotropy but $a^2>0$ case. The smaller radii branch
 corresponds to negative specific heat, which implies a Hawking-Page transition from a black brane  set up to a thermal AdS space. This can be seen clearly from the entropy density-temperature plot given in Fig.\ref{solx} (third) by the top  red line with one branch $\partial s/\partial T<0$ and another $\partial s/\partial T>0$.

 In contrast, for the ``oblate" case with $a^2<0$, as shown in the center graph of Fig.\ref{solx}, there is only one stable black brane configuration and thus is thermodynamically stable. The specific heat is hence positive for all temperatures (see the blue line in Fig.\ref{solx} (third)). At a fixed temperature,
  the horizon radius and the entropy density are less than those of isotropic RN-AdS black brane. The analytical discussion presented here is consistent with the previous numerical result.

 {\textbf{Zero temperature limit} From equation (\ref{tema}), we learn that the black brane temperature can approach zero only when the inverse horizon radius $\uh$ takes the form
\be \label{zeroT}
\uh={4\sqrt{3}[(2-q^2)\sqrt{1+4q^2}]^{1/2}}{\bigg[a^2\bigg(4\sqrt{1+4q^2}-5(2+5q^2)\log\big(\frac{3+\sqrt{1+4q^2}}{3-\sqrt{1+4q^2}}\big)\bigg)\bigg]^{-1/2}}.\nonumber
\ee
The positiveness of the horizon radius  requires that the axion field parameter $a$ must be imaginary-valued.
If the anisotropy constant $a$ takes a real number, equation (\ref{zeroT}) cannot be satisfied for any physical $q$ and $\uh$. This is in agreement with the previous numerical analysis that
``prolate" black brane solution yields no extremal configurations. Fig.\ref{solx} (middle) plots the temperature as a function of the horizon radius. The temperature becomes a monotonic function of the horizon radius and zero temperature is available for imaginary-valued anisotropy.

The appearance of the imaginary valued $a$ can be interpreted as follows: The coupling of the dilatonic field to the axion field induces an effective negative mass term for the dilatonic field. In our case, in the dilatonic equation of motion $\nabla_{\mu}\nabla^{\mu}\phi-e^{2\phi}(\partial\chi)^2=0$,
$e^{2\phi}(\partial\chi)^2$ corresponds to  the mass term. As the temperature is lowered, this mass term eventually drives the dilatonic field tachyonic.  So that the dilatonic field could condensate in the IR, similar to that observed in the condensation of neutral scalar field in holographic superconductors\cite{horowitz}. Note that although the mass squared term is negative, it is above the Breitenlohner-Freedman (BF) bound.


 \section{Holographic stress tensor}
 A concrete calculation on holographic renormalization for this charged system is presented in \cite{cgs} and it is proved that the presence of the $U(1)$ gauge field contributes no additional logarithmic divergences. The counter terms to action (\ref{5action}) are the same as those of axion-dilaton-gravity system discussed in \cite{mateos}
 \be
 S_{ct}=\frac{1}{\kappa^2}\int d^4x \sqrt{\gamma}(3-\frac{1}{8}e^{2\phi}\partial_i\chi\partial^i\chi)-\log v\int d^4x \sqrt{\gamma} \mathcal{A},
 \ee
 where $v$ is the Ferrerman-Graham (FG) coordinate, $\gamma$ is the induced metric on a $v=v_0$ surface.
 A detailed analysis reveals that the $U(1)$ gauge field does not change the value of the conformal anomaly $\mathcal{A}=\langle T^i_i \rangle=\frac{N^2_c a^4}{48 \pi^2}$ obtained in \cite{mateos}. Note that the stress tensor is diagonal $\langle T_{ij}\rangle=diag(E, P_{xy}, P_{xy}, P_z)$ and obeys $\partial^i \langle T_{ij}\rangle=0$.
 Due to the presence of conformal anomaly, the transformation of the stress tensor under a rescaling of $a$, $T$ and $\mu$ contains an inhomogeneous term
 \be
 \langle T_{ij}(ka,kT,k \mu)\rangle=k^4 \langle T_{ij}(a,T,\mu)\rangle+k^4 \log k~\mathcal{A} h_{ij},
 \ee where $h_{ij}=diag(1,-1,-1,3)$. In turn, the stress tensor has the form
  \be
\langle T_{ij}(a,T,\mu)\rangle=a^4 t_{ij}\left(\frac{a}{T},\frac{a}{\mu}\right)+ \log\left(\frac{a}{\Lambda}\right)\frac{\nc^2 a^4}{48 \pi^2} h_{ij},
\ee
where $\Lambda$ is an arbitrary reference scale, a remnant of the renormalization process like the substraction point in QCD. Different choices of $\Lambda$ correspond to different choices of renormalization scheme.
 This means the physics depends on three dimensionless ratio $T/\Lambda$, $a/\Lambda$ and $\mu/\Lambda$. The phase diagram of the thermodynamics is then deeply  influenced by this reference scale $\Lambda$, because the energy density and the pressure are dependent on $\Lambda$ i.e. scheme-dependent. However, we stress that the $U(1)$ chemical potential $\mu$ is scheme-independence. Under a  rescaling of the coordinates of the form
 $ x_{i}=k x'_{i}, ~~v=k v' $ this would not shift $\mu$ and there is no such logarithmic term  as $ \log v$ in the expression for $\mu$. The scheme-independence of $\mu$ is also implied by its thermodynamic definition \cite{cgs}.

 One can see from the original papers \cite{mateos} that the reference scale $\Lambda$ plays a crucial role in the phase diagram. The introduce of a reference scale $\Lambda$ aiming
 to define the theory with $a\neq 0$ is a direct consequence of the conformal anomaly, in analogy with the situation in QCD with one quark flavor with $M_{q}\neq 0$.
The aim of this paper is that even without considering the renormalization scale $\Lambda$, there exists an alternative type of instability.

\section {Thermodynamics and phase structure}
 The grand canonical thermodynamical potential $\Omega=E-Ts-\mu \rho=-P_{xy}$ can be evaluated  from the on-shell Euclidean action and the entropy density satisfies
 $(\frac{\partial \Omega}{\partial T})_{\mu}=-s$ \cite{cgs}. The pressure along the $z$-direction can be evaluated via $P_z=P_{xy}+(\frac{\partial \Omega}{\partial a})a=-G$. Our computations demonstrate that  the thermodynamics variables corresponding to the ``prolate" and ``oblate" solutions are different:
 \begin{center}
~~~~~~~~~{\rm prolate}: ~~$a^2>0$,~~$s>s^0$,~~$\Omega<\Omega^0$,~~$P_z<P^0$,~~$\mu<\mu^0$,\\
~~~~~~~~{\rm ~~oblate}: ~~$a^2<0$,~~$s<s^0$,~~$\Omega>\Omega^0$,~~$P_z>P^0$,~~$\mu>\mu^0$,
\end{center}
  where  we have not included the contribution of $\Lambda$, and $s^0$, $\Omega^0$, $P^0$ and $\mu^0$ denote the entropy density, thermodynamical potential, pressure and chemical potential of the isotropic RN-AdS black brane. $\Omega$ and $P_z$ are $\Lambda$-dependent, but $s$ and $\mu$ are $\Lambda$-independent.

  We emphasize that the presence of the $U(1)$ chemical potential significantly changes the phase structure of the whole system.
   Note that a charged black brane in AdS space can be
   considered as a system with an infinite charge reservoir and the chemical potential eventually equilibrate to the same value everywhere, then the chemical potential at phase equilibrium should be the same in the isotropic and anisotropic regions of QGP.
   For the ``prolate" solution, the chemical potential of the isotropic phase is higher than that of the anisotropic phase. This means that the anisotropic phase is more stable than the isotropic phase. As a consequence, charges or baryons would immigrate from the isotropic phase to the anisotropic phase. While for the ''oblate'' solution, the anisotropic chemical potential is greater than the isotropic case, implying a metastable state of the anisotropic plasma. That is to say,  charges or baryons would escape from the  anisotropic region to the isotropic region.

 The necessary and sufficient condition for local thermodynamic stability are written as
\be
c_{\rho}\equiv T\bigg(\frac{\partial s}{\partial T}\bigg)_{\rho}> 0, ~~~
 \mu'\equiv\bigg(\frac{\partial  \mu}{\partial \rho}\bigg)_{T}>0.\label{chemical}
\ee
The heat capacity $c_{\rho}$ at constant charges $\rho$  should be positive and regular. The second   condition (\ref{chemical})  states that the system is stable against infinitesimal charge  fluctuations. For ``prolate" solution, we have already shown that the specific heat takes a negative value at the smaller horizon radius, while it is positive for ``oblate" solution. Further evidence can be found in the $\mu-\rho$ diagram (see Fig.\ref{murho}).
\begin{figure}[htbp]
 \begin{minipage}{1\hsize}
\begin{center}
\includegraphics*[scale=0.6] {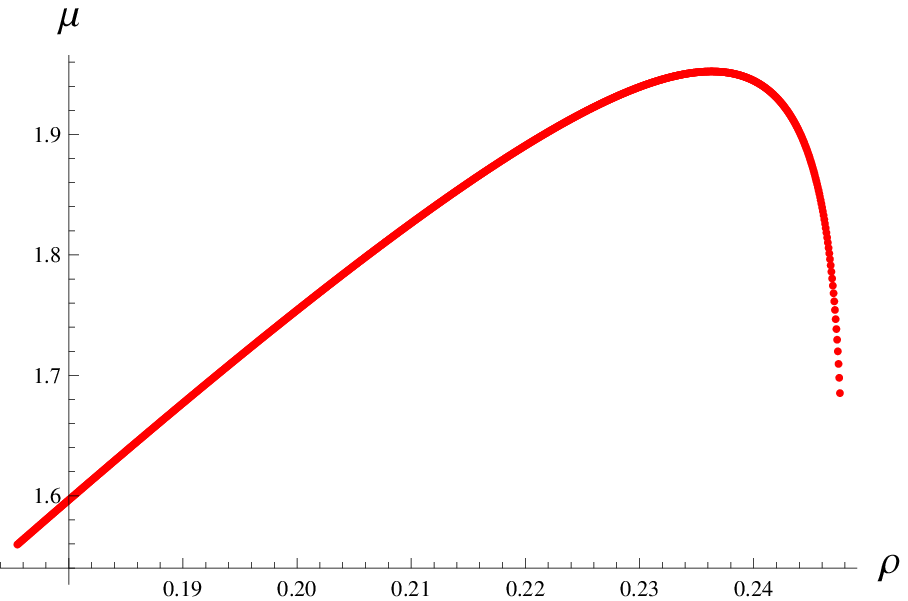}
\includegraphics*[scale=0.6] {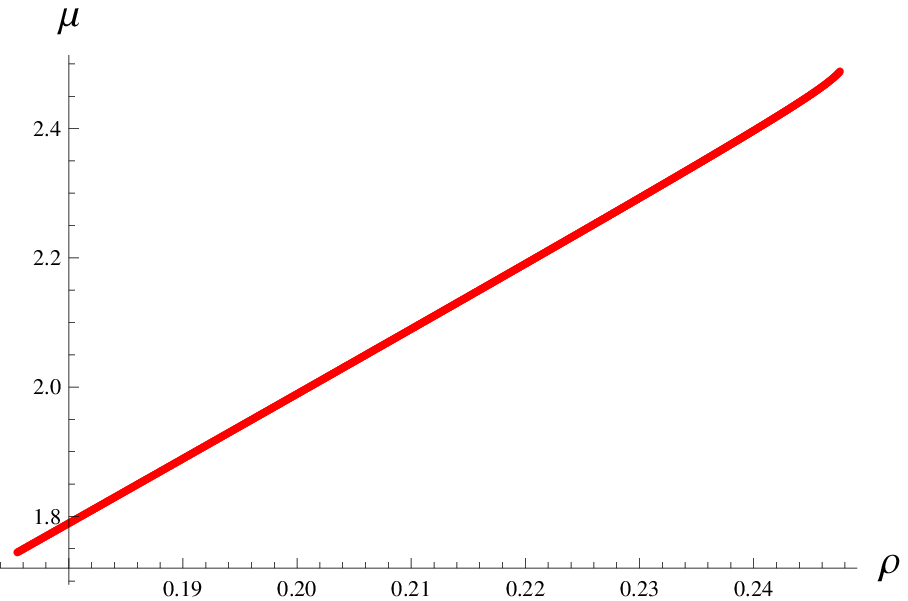}
\end{center}
\caption{ The chemical potential as a function of the charge density for ``prolate" and ``oblate" solutions, respectively. The values $N_c=2$, $\uh=1$ and $a=0.7,0.2i$ have been used here.
The left graph shows that for the ``prolate" solutions, $\partial \mu/\partial\rho<0$ at some regions of $\rho$, signalling an instability of the thermodynamics. On the right, the ``oblate" solution is stable as $\partial \mu/\partial\rho>0$ is satisfied. } \label{murho}
\end{minipage}
\end{figure}

\begin{figure}[htbp]
 \begin{minipage}{1\hsize}
\begin{center}
\includegraphics*[scale=0.6] {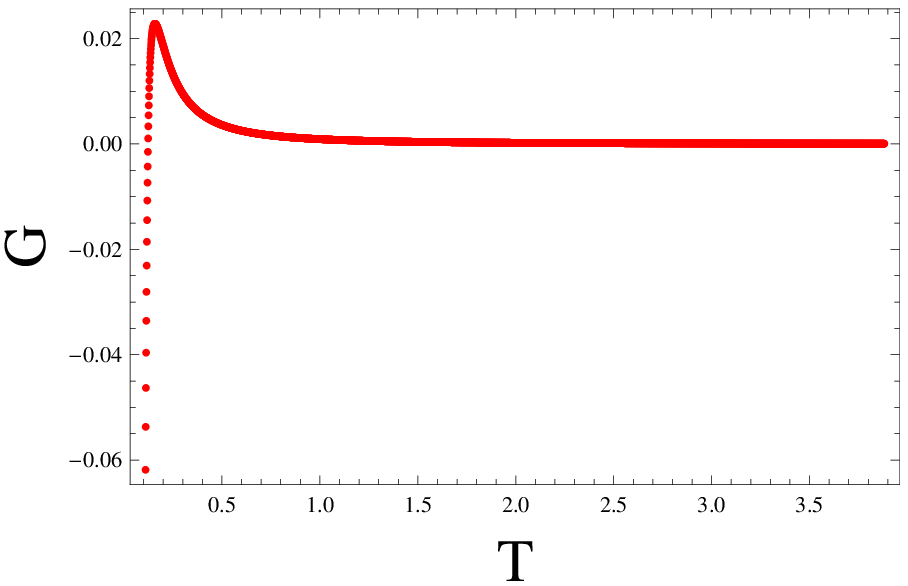}
\includegraphics*[scale=0.62] {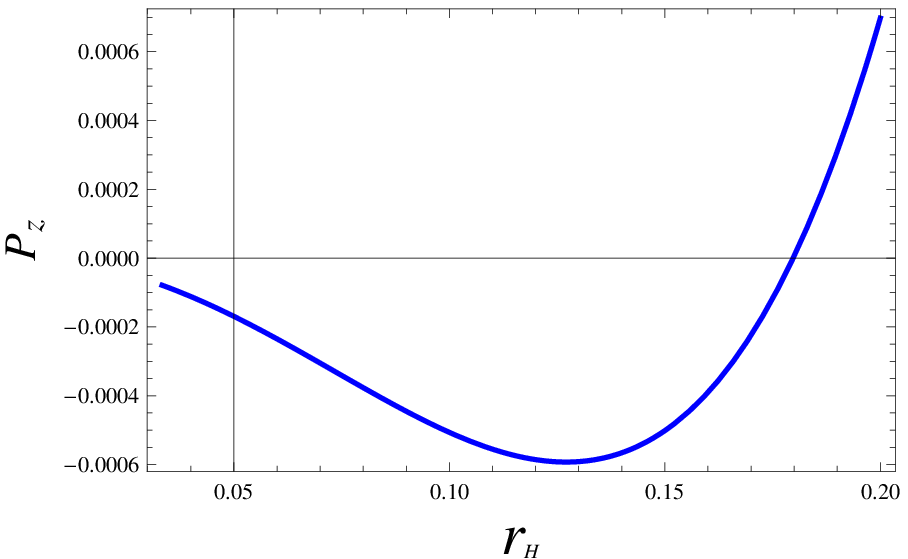}
\end{center}
\caption{(Left) The thermodynamic potential $G$ as a function of temperature for fixed charge and anisotropy, where we set $N_c=8$, $a=0.4$ and $q=1$. (Right) The pressure $P_z$
as a function of horizon radius $r_H$ for fixed charge and anisotropy, where we set $N_c=8$, $a=0.5$ and $q=1.8$. In both case, we neglect contribution of the reference scale $\Lambda$.} \label{GT}
\end{minipage}
\end{figure}
First, consider the ``prolate" black brane solution, the thermodynamic potential $G$ as a function of the temperature has a ``cusp" shape, which is qualitatively the same as the Schwarzschild-AdS black holes.  The thermodynamic potential is positive for some range of $T$, and it is only above the  critical temperature $T_c$ that the thermodynamics is dominated by the black brane phase. This can be seen clearly from Fig.\ref{GT}. The pressure along the $z$-direction is negative at smaller horizon radii. As to the ``oblate" solutions, the thermodynamic potential is strictly negative for all the temperatures and thus is thermodynamically stable.

\section{Conclusion and Discussion}
In the present paper, we have obtained the new charged anisotropic black brane solutions, which might be ``prolate" or ``oblate".
We have mainly compared the thermal properties of these two solutions, and uncovered a new instabilities by using the scheme-independent parameters $s$, $\mu$ and $T$ for the ``prolate" solutions. Generally, for the black holes in equilibrium with the heat bath, the increase in the temperature
  leads to the increase in the black hole radius and mass for stable black holes.
 However, from Fig.\ref{inverseT}(left), we can see that the local slope of the $1/T$ curve is positive for the smaller radii branch, meaning that the temperature decreases as $r_{H}$ increases,  which is quite similar to the familiar case of the uncharged Schwarzschild-AdS black bole with $S^3$ horizon topology.
Therefore, the smaller branch with smaller radii is unstable, having negative specific heat. The smaller branch solution is unphysical and should not be applied to studying the dual CFT.

Note that  the anisotropy constant ``a", the temperature and the horizon radius $r_H$  have the same dimension of mass. The instability uncovered here is due to a competing effect between the scale set by the anisotropy and the scale set by the temperature .
 The instability revealed here is independent of the reference scale $\Lambda$  i.e. scheme-independent.
   If we include the effects of the conformal anomaly $\Lambda$, we can reproduce the similar results discussed in \cite{mateos}.

On the other hand, for the ``oblate" case with $a^2<0$, the specific heat is positive everywhere. In this case, for a given temperature $T$, the horizon radii $r_{H}$ and the entropy density are less than those of  RN-AdS black brane. We also note that $\mu>\mu^0$ in this case. Ignoring the reference scale $\Lambda$,
the pressure along the z-direction satisfies $P_z> P^0$, inferring mechanical stability of the black brane.

 The  potential $\Omega$ reduces to the free energy $F$ when the $U(1)$ charge is absent. In \cite{mateos}, Mateos and Trancanelli have investigated the ``prolate" solution by exploring the scheme-dependent free energy $F$. Specially, through inspecting $F''=(\partial^2 F/\partial a^2)_T$, they clarified the phases into three zones, namely, unstable zones, metastable zones and stable zones determined by the ratio $a/\Lambda$.
 The unstable zones correspond to unstable thermal equilibriums against infinite charge-``a" fluctuations, while the stable zones correspond to metastable thermal equilibriums against finite charge-``a" fluctuations.The unstable and metastable states they uncovered will fall apart into a mixed phase similar to the high-density anisotropic `droplet' or `filaments' surrounded by isotropic regions \cite{mateos}. However, the instabilities uncovered here cannot be rescued by adding the conformal anomaly term, because they are scheme-independent.


{\bf Acknowledgments}\\
We are indebted to Hong Lu and Jian-Xin Lu for help on the Type IIB supergravity action. We would like also to thank LiQing Fang, Xiao-Mei Kuang, Jia-Rui Sun,  Shang-Yu Wu, Yi Yang and Shao-Jun Zhang   for helpful discussions. XHG was supported by NSFC,
China (No.11375110),  and Shanghai Rising-Star Program (No.10QA1402300). SJS was supported by the NRF, Korea (NRF-2013R1A2A2A05004846).

\end{document}